\documentstyle[preprint,aps,eqsecnum]{revtex}
\begin{document}
\draft

\title{\Large\bf{A superspace formulation of \\
Abelian antisymmetric tensor gauge theory}}
\vspace{.3in}
\author{Shinichi Deguchi\footnote{Email address:
deguchi@phys.cst.nihon-u.ac.jp }}
\address{Atomic Energy Research Institute, \\College of Science and
Technology, \\ Nihon University,  Tokyo 101-8308, Japan
 and}
\author{Bhabani Prasad Mandal\footnote{Email address:
bpm@tnp.saha.ernet.in}}
\address{Theory Group,\\ Saha Institute of Nuclear Physics,\\ 1/AF
Bidhannagar, Calcutta- 700 064, India}
\vspace{.3in}

\maketitle

\begin{abstract}
We apply a superspace formulation to the four-dimensional gauge theory 
of a massless Abelian antisymmetric tensor field of rank 2.
The theory is formulated in a six-dimensional superspace using
rank-2 tensor, vector and scalar superfields 
and their associated supersources.
It is shown that BRS transformation rules of fields are realized 
as Euler-Lagrange equations without assuming the so-called 
horizontality condition and that a generating functional $\bar{W}$ 
constracted in the superspace reduces to that for the ordinary 
gauge theory of Abelian rank-2 antisymmetric tensor field.
The WT identity for this theory is derived by making use of 
the superspace formulation 
and is expressed in a neat and compact form 
$\partial\bar{W}/\partial\theta=0$. 
\end{abstract}

\newpage
\section{INTRODUCTION}
Gauge theories of Abelian rank-2 antisymmetric tensor fields have
become field of interest for various reasons.
Kalb and Ramond first realized that Abelian rank-2 antisymmetric
tensor fields could interact with classical strings \cite{kalb}.
This interaction has been applied to the Lorentz-covariant description
of vortex motion in an irrotational, incompressible fluid \cite{lund},
and to the dual formulation of the Abelian Higgs model \cite{suga}.
Abelian rank-2 antisymmetric tensor fields are also involved in
supergravity multiplets \cite{sgr} and in excited states of
quantized (super)strings \cite{str}.
They are crucial for superstring theories to
realize anomaly-cancellation mechanism and to estimate dualities
in extended objects.
In addition, it has been shown that an Abelian rank-2 antisymmetric
tensor field generates effective mass for an Abelian vector gauge field
through a topological coupling between these two fields \cite{top}.
A geometric aspect of Abelian rank-2 antisymmetric tensor
fields has been discussed in a U(1) gauge theory in loop space
\cite{loop}.

Covariant quantization of an Abelian rank-2 antisymmetric tensor field
was first attempted by Townsend \cite{town} and has been studied by
many authors [9,10] in systematic manners based on
the Becchi-Rouet-Stora (BRS) formalism.
It was found in the covariant quantization that a naive
gauge-fixing term containing the antisymmetric tensor field
is itself invariant under a secondary gauge transformation and
that commuting ghost fields are required for complete gauge fixing.

In the present paper, we consider a superspace formulation of the
four-dimensional gauge theory of a massless Abelian rank-2
antisymmetric tensor field.
Until recently, the BRS formalism for Abelian rank-2 antisymmetric
tensor fields has been discussed by several authors using
superfields on a six-dimensional superspace \cite{supe}.
However, in these superspace formulations,
BRS transformation rules of fields are put in
by hand and superfields are not free but are constrained to satisfy
the so-called horizontality condition.
As a consequence, covariance of the superfields is spoiled
under superspace rotations that mix spacetime and anticommuting
coordinates.

To avoid these limitations,
we apply the superspace formulation
proposed by Joglekar \cite{sdj} to the gauge theory of
Abelian rank-2 antisymmetric tensor field.
This formulation also uses a six-dimensional superspace but
has remarkable features:
(i) Unlike earlier superspace formulations [10,12],
superfields are not a priori restricted by the horizontality
conditions and superspace rotations can be carried out;
BRS transformation rules are realized as
Euler-Lagrange equations.
(ii) The theory in superspace is constructed by
taking into account
generalized gauge invariance and the Lagrangian density
(without gauge-fixing and source terms)
is a scalar under $ {\rm OSp}(3,1|2) $ transformations in
the superspace.
(iii) The whole action including all source terms
is accommodated in a single action written in terms of superfields.
For details see Ref. \cite{bpm}.
After the application, we show that the generating functional
in the superspace formulation contains in itself all
the necessary information of the generating functional
for the ordinary gauge theory of
Abelian rank-2 antisymmetric tensor field.
We further show that BRS transformation rules
can be obtained without assuming the horizontality
condition and that the WT identity for the theory
can be expressed in a neat, compact, and
mathematically convenient form $\partial\bar{W}/\partial\theta=0$.
This will lead to a simplified treatment of the renormalization
problem in the gauge theory of Abelian rank-2 antisymmetric
tensor field.

The present paper is organized as follows.
We briefly review the gauge theory of Abelian
rank-2 antisymmetric tensor field in Sec.IIA
and provide for the superspace formulation in Sec.IIB.
In Sec.III, we apply the superspace formulation proposed
in Ref.\cite{sdj} to
the gauge theory of Abelian rank-2 antisymmetric tensor field
and construct a generating functional from superfields.
We show that this generating functional reduces to one
considered in the ordinary gauge theory of
Abelian rank-2 antisymmetric tensor field.
In Sec.IV, we see that BRS transformation rules are
realized as Euler-Lagrange equations.
We also discuss a relation between
the BRS transformation and a six-dimensional gauge transformation.
In Sec.V, we derive a simple form of the WT identity by making use of
the superfield formulation.
Section VI is devoted to a summary and discussion.

\section{PRELIMINARY}
\subsection{Gauge theory of Abelian rank-2 antisymmetric tensor field}
In this section, we briefly review the gauge theory
of Abelian rank-2 antisymmetric tensor field.
Let $B_{\mu\nu}(x)$ be an Abelian antisymmetric tensor field on
four-dimensional Minkowski space $M^{4}$ with a space-time coordinate system
$(x^{\mu})$ $(\mu=0,\,1,\,2,\,3)$. We consider
the Abelian gauge theory defined by the action
\begin{equation}
S_{0}={1\over12} \int d^{4}x F_{\mu\nu\rho}F^{\mu\nu\rho}  ,
\label{1}
\end{equation}
%
where $d^{4}x\equiv dx^{0}dx^{1}dx^{2}dx^{3}$ and
$F_{\mu\nu\rho}\equiv \partial_{\mu}B_{\nu\rho}+
\partial_{\nu}B_{\rho\mu}+\partial_{\rho}B_{\mu\nu}$.
This action is invariant under the gauge transformation
$\delta B_{\mu\nu}=\partial_{\mu}\Lambda_{\nu}
-\partial_{\nu}\Lambda_{\mu}$ with a vector gauge parameter
$\Lambda_{\mu}(x)$.

To covariantly quantize $B_{\mu\nu}$ using the BRS formalism [9,10],
it is necessary to introduce the following ghost and auxiliary
fields: anticommuting vector fields
$\rho_{\mu}(x)$ and $\tilde{\rho}_{\mu}(x)$,
a commuting vector field $\beta_{\mu}(x)$,
anticommuting scalar fields $\chi(x)$ and $\tilde{\chi}(x)$,
and commuting scalar fields $\sigma(x)$, $\varphi(x)$ and
$\tilde{\sigma}(x)$.
The BRS transformation $\mbox{\boldmath{$\delta$}}$ is defined for
$B_{\mu\nu}$ by replacing $\Lambda_{\mu}$ in
the gauge transformation by the ghost field $\rho_{\mu}$,
and for other fields it is defined so as to
satisfy the nilpotency condition $\mbox{\boldmath{$\delta$}}^{2}=0\,$:
\begin{eqnarray}
\mbox{\boldmath{$\delta$}} B_{\mu\nu} &=&
\partial_{\mu}\rho_{\nu}-\partial_{\nu}\rho_{\mu}
\nonumber \\
\mbox{\boldmath{$\delta$}}\rho_{\mu}&=&-i\partial_{\mu}\sigma \,,
\quad
\mbox{\boldmath{$\delta$}}\sigma=0 \,,
\nonumber \\
\mbox{\boldmath{$\delta$}}\tilde{\rho}_{\mu}&=&i\beta_{\mu} \,,
\qquad
\mbox{\boldmath{$\delta$}}\beta_{\mu}=0 \,,
\nonumber \\
\mbox{\boldmath{$\delta$}}\tilde{\sigma}&=&\tilde{\chi} \,,
\qquad \quad \,
\mbox{\boldmath{$\delta$}}\tilde{\chi}=0 \,,
\nonumber \\
\mbox{\boldmath{$\delta$}}\varphi&=&\chi \,,
\qquad \quad \,
\mbox{\boldmath{$\delta$}}\chi=0 \:.
\label{2}
\end{eqnarray}
%
Covariant quantization of $B_{\mu\nu}$ can be performed with the action
\begin{equation}
S=S_{0}+S_{1}+S_{2} \,,
\label{3}
\end{equation}
%
with the gauge-fixing terms
\begin{eqnarray}
S_{1}&=& -i\int d^{4}x \,
\mbox{\boldmath{$\delta$}}
[\tilde{\rho}_{\nu} (\partial_{\mu}B^{\mu\nu}
+k_{1}\beta^{\nu} )] \, ,
\label{4}
\\
S_{2}&=& -i\int d^{4}x \,
\mbox{\boldmath{$\delta$}}
[\tilde{\sigma} \partial_{\mu}\rho^{\mu}
+\varphi ( \partial_{\mu}\tilde{\rho}^{\mu}
-k_{2}\tilde{\chi} ) ] \,,
\label{5}
\end{eqnarray}
%
where $k_{1}$ and $k_{2}$ are gauge parameters.
Owing to the nilpotency property of $\mbox{\boldmath{$\delta$}}$,
these gauge-fixing terms are invariant under the BRS transformation.
The first term $S_{1}$ breaks the gauge invariance
of $S_{0}$ explicitly.
The second term $S_{2}$ is necessary to break
the invariance of $S_{1}$ under the secondary gauge
transformation $\delta\rho_{\mu}=\partial_{\mu}\varepsilon, \;
\delta\tilde{\rho}_{\mu}=\partial_{\mu}\tilde{\varepsilon}$.
The gauge-fixing procedure for quantization of $B_{\mu\nu}$ is
complete with $S_{1}+S_{2}$.
Carrying out the BRS transformation in Eqs. (\ref{4}) and (\ref{5}), we
obtain
\begin{eqnarray}
S_{1}+S_{2}&=&
\int d^{4}x [
-i\partial_{\mu}\tilde{\rho}_{\nu}
(\partial^{\mu}\!\rho^{\nu}-\partial^{\nu}\!\rho^{\mu})
+\partial_{\mu}\tilde{\sigma}\partial^{\mu}\sigma
\nonumber \\
& &\;+\beta_{\nu} (\partial_{\mu}B^{\mu\nu}
+k_{1}\beta^{\nu}-\partial^{\nu}\varphi)
\nonumber \\
& &\;-i\tilde{\chi}\partial_{\mu}\rho^{\mu}
-i\chi(\partial_{\mu}\tilde{\rho}^{\mu}
-k_{2}\tilde{\chi}) ] \,.
\label{6}
\end{eqnarray}
%

The action $S$ describes a massless system.
We can read from $S$ how many physical degrees of freedom
$B_{\mu\nu}$ has: The total degrees of freedom of the commuting fields
$B_{\mu\nu}$, $\beta_{\mu}$, $\sigma$, $\varphi$ and $\tilde{\sigma}$
are naively 13, but some of them are not independent because of the
four constraints
$\partial_{\mu}B^{\mu\nu}+k_{1}\beta^{\nu}-\partial^{\nu}\varphi=0$
derived from $S$. Their genuine degrees of freedom are thus 9.
The total degrees of freedom of the anticommuting fields $\rho_{\mu}$
$\tilde{\rho}_{\mu}$, $\chi$ and $\tilde{\chi}$ are naively 10, but
some of them are also not independent because of the two constrains
$\partial_{\mu}\rho^{\mu}=0$ and
$\partial_{\mu}\tilde{\rho}^{\mu}-k_{2}\tilde{\chi}=0$
derived from $S$. Their genuine degrees of freedom are thus 8.
Subtracting the genuine degrees of freedom of the anticommuting fields
from those of the commuting fields, we conclude that $B_{\mu\nu}$ has
one physical degree of freedom, describing a spin less particle.

\subsection{ Superspace and superfields }
We shall work in a superspace of six dimensions.
The superspace used in this paper possesses a local coordinate system
$(\bar{x}^{i})\equiv (x^\mu, \lambda , \theta )$
$(i=0,\,1,\,2,\,3,\,4,\,5)$ with anticommuting real coordinates
$x^4\equiv\lambda$ and $x^5\equiv\theta$, and will be denoted by
$M^{4/2}$. We introduce to $M^{4/2}$ a metric tensor $g_{ij}$
whose non-vanishing components are
\begin{equation}
g_{00} = -g_{11} = -g_{22} = -g_{33} =-g_{45} = g_{54} =1 \,.
\label{7}
\end{equation}
%
The set of linear homogeneous transformations that leave
$g_{ij}\bar{x}^i\bar{x}^j$ invariant forms the pseudo-orthosymplectic
supergroup ${\rm OSp}(3,1|2)$. This is nothing but a supersymmetric
generalization of the Lorentz group.

Let $\bar{X}(\bar{x})=\bar{X}(x,\lambda,\theta)$ be an arbitrary
superfield on $M^{4/2}$.
\footnote[1]{In this paper we shall attach
^^ ^^ overbar" to {\it all} the superfields on $M^{4/2}$.}
Since $\lambda$ and $\theta$ are nilpotent, $\bar{X}$ can be expanded
as
\begin{equation}
\bar{X}(\bar{x})=X(x)+\lambda X_{\lambda}(x)+\theta X_{\theta}(x)
+\lambda\theta X_{\lambda\theta}(x) \, ,
\label{8}
\end{equation}
%
where $X$, $X_{\lambda}$, $X_{\theta}$ and $X_{\lambda\theta}$
are component fields on $M^{4}$. In terms of
\begin{eqnarray}
\bar{X}_{,\lambda}(\bar{x}) &\equiv&
{{\partial\bar{X}(\bar{x})}\over{\partial\lambda}}
=X_{\lambda}(x)+\theta X_{\lambda\theta}(x) \,,
\nonumber \\
\bar{X}_{,\theta}(\bar{x}) &\equiv&
{{\partial\bar{X}(\bar{x})}\over{\partial\theta}}
=X_{\theta}(x)-\lambda X_{\lambda\theta}(x) \,,
\label{9}
\end{eqnarray}
%
and $X_{\lambda\theta}(x)$, Eq. (\ref{8}) is written as
\begin{equation}
\bar{X}(\bar{x})=X(x)+\lambda \bar{X}_{,\lambda}(\bar{x})
+\theta \bar{X}_{,\theta}(\bar{x})
-\lambda\theta X_{\lambda\theta}(x) \, .
\label{10}
\end{equation}
%
Here ${\partial\over{\partial\lambda}}$
and ${\partial\over{\partial\theta}}$ denote left derivatives.
Equation (\ref{10}) can be regarded as a constraint in the five fields
$\bar{X}$, $X$, $\bar{X}_{,\lambda}$, $\bar{X}_{,\theta}$ and
$X_{\lambda\theta\,}$; we now choose
${\cal X}\equiv(\bar{X},\, \bar{X}_{,\lambda},\,
\bar{X}_{,\theta},\, X_{\lambda\theta})$ as a set of independent
fields.

Let us define
\begin{equation}
\bar{f}
\equiv\int d^{4}x f\!\left( {\cal X}(\bar{x}),\,
\partial_{\mu}{\cal X}(\bar{x}) \right)
\label{11}
\end{equation}
%
from a polynomial $f$ in ${\cal X}$ and $\partial_{\mu}{\cal X}$.
For an arbitrary function ${\cal F}$ of $\bar{f}$,
we can readily show that
\begin{eqnarray}
& &\int \prod_{x} d\bar{X}(\bar{x}) d\bar{X}_{,\lambda}(\bar{x})
d\bar{X}_{,\theta}(\bar{x}) dX_{\lambda\theta}(x)
{\cal F}
\nonumber \\
& & \; =\int \prod_{x} dX(x) dX_{\lambda}(x) dX_{\theta}(x)
dX_{\lambda\theta}(x) {\cal F} \,.
\label{12}
\end{eqnarray}
%
Equation (\ref{12}) holds whether $\bar{X}$ is a commuting
superfield or an anticommuting superfield. It should be noted that
the integration in Eq. (\ref{11}) and the multiplication in Eq. (\ref{12})
are
carried out over $M^{4}$, not over $M^{4/2}$.
Consequently, Eq. (\ref{12}) is still a function of $\lambda$
and $\theta$.
If $f$ does not depend on $X_{\lambda\theta}$ and
$\partial_{\mu}X_{\lambda\theta}$,
we can formally divide the both sides of Eq. (\ref{12})
by $\int \prod_{x} dX_{\lambda\theta}(x)$, arriving at
\begin{eqnarray}
& &\int \prod_{x} d\bar{X}(\bar{x}) d\bar{X}_{,\lambda}(\bar{x})
d\bar{X}_{,\theta}(\bar{x}) {\cal F}
\nonumber \\
& & \; =\int \prod_{x} dX(x) dX_{\lambda}(x) dX_{\theta}(x)
{\cal F} \,.
\label{13}
\end{eqnarray}
%
Using this formula twice, we have
\begin{eqnarray}
& &{\partial\over{\partial \theta}}
\int \prod_{x} d\bar{X}(\bar{x}) d\bar{X}_{,\lambda}(\bar{x})
d\bar{X}_{,\theta}(\bar{x}) {\cal F}
\nonumber \\
& & \; =
\int \prod_{x} d\bar{X}(\bar{x}) d\bar{X}_{,\lambda}(\bar{x})
d\bar{X}_{,\theta}(\bar{x})
{{\partial {\cal F}}\over{\partial\theta}} \,.
\label{14}
\end{eqnarray}
%
This is an important formula used in the superspace formulation
of gauge theories.

\section{Action and generating functional in superspace}
In this section, we apply the superfield formulation
proposed by Joglekar \cite{sdj} to the
gauge theory of Abelian rank-2 antisymmetric tensor field.
To this end,
we now generalize the antisymmetric tensor field $B_{\mu\nu}(x)$
to a superfield $\bar{B}_{ij}(\bar{x})$ on $M^{4/2}$
satisfying the antisymmetric property in superspace,
$\bar{B}_{ij}=-(-1)^{|i||j|}\bar{B}_{ji}$,
and the commuting property
$\bar{x}^{k}\bar{B}_{ij}=(-1)^{|k|(|i|+|j|)}\bar{B}_{ij}\bar{x}^{k}$.
Here $|i|$ is a function of $i$, defined as $|i|=0$ for
$i=0,\,1,\,2,\,3$, and $|i|=1$ for $i=4,\,5$.
The superfield $\bar{B}_{ij}$ is assumed to transform
as a rank-2 covariant tensor under
coordinate transformations characterized by ${\rm OSp}(3,1|2)$.
The field strength of $\bar{B}_{ij}$ is defined by
\begin{eqnarray}
\bar{F}_{ijk} \equiv \partial_{i}\bar{B}_{jk}
+(-1)^{|i|(|j|+|k|)}\partial_{j}\bar{B}_{ki}
+(-1)^{|k|(|i|+|j|)} \partial_{k}\bar{B}_{ij} \,,
\label{15}
\end{eqnarray}
%
which is invariant under the generalized gauge transformation
\begin{equation}
\delta\bar{B}_{ij}=\partial_{i}\bar{\Lambda}_{j}-(-1)^{|i||j|}
\partial_{j}\bar{\Lambda}_{i}
\label{16}
\end{equation}
with a vector gauge parameter $\bar{\Lambda}_{i}(\bar{x})$
satisfying the commuting property
$\bar{x}^{k}\bar{\Lambda}_{i}=(-1)^{|k||i|}\bar{\Lambda}_{i}
\bar{x}^{k}$.
We consider the following generalization of the action (\ref{1}):
\begin{equation}
\bar{S}_{0}=-{1\over12} \int d^{4}x \bar{F}^{ijk}(\bar{x})
\bar{F}_{kji}(\bar{x}) \,.
\label{17}
\end{equation}
%
Note that the integration above is carried out over $M^{4}$.
Obviously, $\bar{S}_{0}$ is invariant under
the gauge transformation (\ref{16}).

In addition to $\bar{B}_{ij}$, we introduce a vector superfield
$\bar{\zeta}_{i}(\bar{x})$ satisfying the anticommuting property
$\bar{x}^{k}\bar{\zeta}_{i}
=(-1)^{|k|(|i|+1)}\bar{\zeta}_{i}\bar{x}^{k}$ and
a commuting scalar superfield $\bar{\eta}(\bar{x})$.
Furthermore, we introduce supersources (source superfields)
$\bar{K}^{ij}(\bar{x})$, $\bar{t}^{i}(\bar{x})$ and
$\bar{u}(\bar{x})$ which are associated with the
superfields $\bar{B}_{ij}$, $\bar{\zeta}_{i}$ and $\bar{\eta}$,
respectively. It is assumed that the (inner) products
$\bar{K}^{ij}\bar{B}_{ji}$, $\bar{\zeta}_{i}\bar{t}^{i}$
and $\bar{\eta}\bar{u}$ are anticommuting scalars under
${\rm OSp}(3,1|2)$ transformations; that is,
$\bar{K}^{ij}$ is an anticommuting tensor supersource,
$\bar{t}^{i}$ a commuting vector supersource, and
$\bar{u}$ an anticommuting scalar supersource.

The superspace formulation of the gauge theory of Abelian rank-2
antisymmetric tensor field is begun with the action
\begin{equation}
\bar{S}=\bar{S}_{0}+\bar{S}_{\rm GS}
\label{18}
\end{equation}
%
with the gauge-fixing and source terms
\begin{eqnarray}
\bar{S}_{\rm GS}&=&\int d^{4}x \,{\partial\over{\partial\theta}}
\Biggl[ \bar{K}^{ij}(\bar{x})\bar{B}_{ji}(\bar{x}) \Biggr.
\nonumber \\
& & \; +\bar{\zeta}_{i}(\bar{x})
\{ {\delta^{i}}_{\alpha}
\partial_{\mu}\bar{B}^{\mu\alpha}(\bar{x})
+k_{1}{\delta^{i}}_{\nu}
\bar{\zeta}\raisebox{0.95ex}{\mbox{\scriptsize{$\nu$}}}
\raisebox{-0.36ex}{\mbox{\scriptsize{$,\!\theta$}}}(\bar{x})
+\bar{t}^{i}(\bar{x}) \}
\nonumber \\
& & \; \Biggl. +\bar{\eta}(\bar{x})
\{ \partial_{\mu}\bar{\zeta}^{\mu}(\bar{x})
+k_{2}\bar{\zeta}_{4,\theta}(\bar{x})
+\bar{u}(\bar{x}) \} \! \Biggr] \,,
\label{19}
\end{eqnarray}
%
where $\alpha=0,\,1,\,2,\,3,\,4$, and $k_{1}$ and $k_{2}$
are gauge parameters.

Let us collectively denote the superfields by $\bar{\Phi}(\bar{x})$
and the supersources by $\bar{\Sigma}(\bar{x})$:
$\bar{\Phi}=(\bar{B}_{ij},\,\bar{\zeta}_{i},\,\bar{\eta})$ and
$\bar{\Sigma}=(\bar{K}^{ij},\,\bar{t}^{i},\,\bar{u})$.
In accordance with the discussion in Sec.2.2, we treat
$\bar{\Phi}(\bar{x})$, $\bar{\Phi}_{,\lambda}(\bar{x})$
and $\bar{\Phi}_{,\theta}(\bar{x})$ as a set of independent fields,
and $\bar{\Sigma}(\bar{x})$ and $\bar{\Sigma}_{,\theta}(\bar{x})$
as a set of independent sources.
(The field $\Phi_{\lambda\theta}(x)$
and the sources $\bar{\Sigma}_{,\lambda}(\bar{x})$ and
$\Sigma_{\lambda\theta}(x)$ do not occur at this stage.)
Now, defining the path-integral measure
\begin{equation}
\{ {\cal D}\bar{\Phi} \}\equiv {\cal D}\bar{\Phi}
{\cal D}\bar{\Phi}_{,\lambda} {\cal D}\bar{\Phi}_{,\theta} \;,
\label{20}
\end{equation}
%
with
\begin{eqnarray}
{\cal D}\bar{\Phi} &\equiv& \prod_{x} d\bar{\Phi}(\bar{x}) \,,
\nonumber \\
{\cal D}\bar{\Phi}_{,\lambda} &\equiv&
\prod_{x} d\bar{\Phi}_{,\lambda}(\bar{x}) \,,
\quad
{\cal D}\bar{\Phi}_{,\theta} \equiv
\prod_{x} d\bar{\Phi}_{,\theta}(\bar{x}) \,,
\label{21}
\end{eqnarray}
%
we consider the generating functional
\begin{equation}
\bar{W} [\,\bar{\Sigma}, \bar{\Sigma}_{,\theta\,};
\lambda, \theta\,]
=\int \{ {\cal D}\bar{B}_{ij} \}
\{ {\cal D}\bar{\zeta}_{i} \}
\{ {\cal D}\bar{\eta} \}
\exp(i\bar{S}) \,.
\label{22}
\end{equation}
%
Since the integrations in Eqs. (\ref{17}) and (\ref{19}) and the
multiplications in Eq. (\ref{21}) are carried out over $M^{4}$,
the generating functional $\bar{W}$
should be understood to be a function of $\theta$ and $\lambda$
as well as a functional of $\Sigma$ and $\Sigma_{,\theta}$.

The integrations over $\bar{B}_{ij,\lambda}$ and
$\bar{B}_{ij,\theta}$ in Eq. (\ref{22}) lead to a form of $\bar{W}$
that is proportional to
$\prod_{i,x}\delta(\bar{K}^{4i}(\bar{x}))$.
Then, carrying out the integrations over
$\bar{B}_{i4}$, $\bar{\zeta}_{5}$,
$\bar{\zeta}_{i,\lambda}$, $\bar{\zeta}_{5,\theta}$, and
$\bar{\eta}_{,\lambda}$,
we finally arrive at
\begin{eqnarray}
\bar{W} &=& N\prod_{i,x} \delta(\bar{K}^{4i}(\bar{x}))
\delta(\bar{K}\raisebox{0.97ex}{\mbox{\scriptsize{$4i$}}}
\raisebox{-0.36ex}{\mbox{\scriptsize{$,\!\theta$}}}(\bar{x}))
\prod_{x} \delta(\bar{t}_{4}(\bar{x}))
\delta(\bar{t}_{4,\theta}(\bar{x}))
\nonumber \\
& & \; \times
\int {\cal D}\bar{\cal M} \exp(i\bar{S}^\prime) \,,
\label{23}
\end{eqnarray}
%
where $N$ is a constant,
\begin{equation}
{\cal D}\bar{\cal M}\equiv
{\cal D}\bar{B}_{\mu\nu}
{\cal D}\bar{B}_{\mu5}
{\cal D}\bar{B}_{55}
{\cal D}\bar{\zeta}_{\mu} {\cal D}\bar{\zeta}_{4}
{\cal D}\bar{\zeta}_{\mu,\theta} {\cal D}\bar{\zeta}_{4,\theta}
{\cal D}\bar{\eta} {\cal D}\bar{\eta}_{,\theta} \,,
\label{24}
\end{equation}
and
\begin{equation}
\bar{S}^\prime=\bar{S}'_{0}+\bar{S}_{1,2}+\bar{S}_{\Sigma} \,,
\label{25}
\end{equation}
%
with
\begin{eqnarray}
\bar{S}'_{0}&=&{1\over12} \int d^{4}x
\bar{F}_{\mu\nu\rho}\bar{F}^{\mu\nu\rho} \,,
\label{26}
\\
\bar{S}_{1,2}&=& \int d^{4}x
\Biggl[ -\partial_{\mu}\bar{\zeta}_{\nu}
(\partial^{\mu}\bar{B}\raisebox{0.95ex}{\mbox{\scriptsize{$\nu$}}}
\raisebox{-0.36ex}{\mbox{\scriptsize{5}}}
-\partial^{\nu}\bar{B}\raisebox{0.95ex}{\mbox{\scriptsize{$\mu$}}}
\raisebox{-0.36ex}{\mbox{\scriptsize{5}}})
-{1\over2}\partial_{\mu}\bar{\zeta}_{4}
\partial^{\mu\!}\bar{B}_{55} \Biggr.
\nonumber \\
& & \;
+\bar{\zeta}_{\nu,\theta}(\partial_{\mu} \bar{B}^{\mu\nu}
+k_{1}\bar{\zeta}\raisebox{0.95ex}{\mbox{\scriptsize{$\nu$}}}
\raisebox{-0.36ex}{\mbox{\scriptsize{$,\!\theta$}}}
-\partial^{\nu}\bar{\eta})
\nonumber \\
& & \; \Biggl.
+\bar{\zeta}_{4,\theta}\partial_{\mu}
\bar{B}\raisebox{0.95ex}{\mbox{\scriptsize{$\mu$}}}
\raisebox{-0.36ex}{\mbox{\scriptsize{5}}}
+\bar{\eta}_{,\theta}(\partial_{\mu}\bar{\zeta}^{\mu}
+k_{2}\bar{\zeta}_{4,\theta})
\Biggr] \,,
\label{27}
\\
\bar{S}_{\Sigma}&=& \int d^{4}x
\Bigl[ -\bar{K}^{\mu\nu}
(\partial_{\mu}\bar{B}_{\nu5}-\partial_{\nu}\bar{B}_{\mu5})
-\bar{K}^{\mu5}\partial_{\mu}\bar{B}_{55}
\Bigr.
\nonumber \\
& &\; -\bar{K}\raisebox{0.96ex}{\mbox{\scriptsize{$\mu\nu$}}}
\raisebox{-0.36ex}{\mbox{\scriptsize{$,\!\theta$}}}
\bar{B}_{\mu\nu}
-2\bar{K}\raisebox{0.96ex}{\mbox{\scriptsize{$\mu5$}}}
\raisebox{-0.36ex}{\mbox{\scriptsize{$,\!\theta$}}}
\bar{B}_{\mu5}
+\bar{K}\raisebox{0.96ex}{\mbox{\scriptsize{$55$}}}
\raisebox{-0.36ex}{\mbox{\scriptsize{$,\!\theta$}}}
\bar{B}_{55}
\nonumber \\
& &\;+\bar{t}^{\mu}\bar{\zeta}_{\mu,\theta}
-\bar{t}_{5}\bar{\zeta}_{4,\theta}
+\bar{t}\raisebox{0.95ex}{\mbox{\scriptsize{$\mu$}}}
\raisebox{-0.36ex}{\mbox{\scriptsize{$,\!\theta$}}}\bar{\zeta}_{\mu}
+\bar{t}_{5,\theta}\bar{\zeta}_{4}
\nonumber \\
& &\;
\Bigl.
-\bar{u}\bar{\eta}_{,\theta}+\bar{u}_{,\theta}\bar{\eta} \Bigr] \;.
\label{28}
\end{eqnarray}
We also consider the generating functional $W$ defined by
\begin{eqnarray}
& & W[\bar{K}^{\mu\nu}, \bar{K}^{\mu5},
\bar{K}\raisebox{0.96ex}{\mbox{\scriptsize{$\mu\nu$}}}
\raisebox{-0.36ex}{\mbox{\scriptsize{$,\!\theta$}}},
\bar{K}\raisebox{0.96ex}{\mbox{\scriptsize{$\mu5$}}}
\raisebox{-0.36ex}{\mbox{\scriptsize{$,\!\theta$}}},
\bar{K}\raisebox{0.96ex}{\mbox{\scriptsize{$55$}}}
\raisebox{-0.36ex}{\mbox{\scriptsize{$,\!\theta$}}},
\bar{t}^{\mu}, \bar{t}_{5},
\bar{t}\raisebox{0.95ex}{\mbox{\scriptsize{$\mu$}}}
\raisebox{-0.36ex}{\mbox{\scriptsize{$,\!\theta$}}},
\bar{t}_{5,\theta},
\bar{u}, \bar{u}_{,\theta\,}; \lambda, \theta\,] \;\;\;\;
\nonumber \\
& & \; =\int {\cal D}\bar{\cal M} \exp(i\bar{S}^\prime) \,,
\label{29}
\end{eqnarray}
%
with which $\bar{W}$ is written as
\begin{equation}
\bar{W} = N\prod_{i,x} \delta(\bar{K}^{4i}(\bar{x}))
\delta(\bar{K}\raisebox{0.97ex}{\mbox{\scriptsize{$4i$}}}
\raisebox{-0.36ex}{\mbox{\scriptsize{$,\!\theta$}}}(\bar{x}))
\prod_{x} \delta(\bar{t}_{4}(\bar{x}))
\delta(\bar{t}_{4,\theta}(\bar{x})) W \,.
\label{30}
\end{equation}
%
Integrating Eq. (\ref{30}) over $\bar{K}^{4i}$,
$\bar{K}\raisebox{0.97ex}{\mbox{\scriptsize{$4i$}}}
\raisebox{-0.36ex}{\mbox{\scriptsize{$,\!\theta$}}}$,
$\bar{t}_{4}$ and $\bar{t}_{4,\theta}$, we have
\begin{equation}
W = \int {\cal D}\bar{K}^{4i}
{\cal D}\bar{K}\raisebox{0.97ex}{\mbox{\scriptsize{$4i$}}}
\raisebox{-0.36ex}{\mbox{\scriptsize{$,\!\theta$}}}
{\cal D}\bar{t}_{4} {\cal D}\bar{t}_{4,\theta} \bar{W} \,.
\label{31}
\end{equation}

With the identification
\begin{eqnarray}
\bar{B}_{\mu\nu}&=&B_{\mu\nu} \,,
\nonumber \\
\bar{B}_{\mu5}&=&i\rho_{\mu} \,,
\quad \;  \bar{B}_{55}=-2i\sigma \,,
\nonumber \\
\bar{\zeta}_{\mu}&=&\tilde{\rho}_{\mu} \,,
\quad \;\:\,
\bar{\zeta}_{\mu,\theta}=\beta_{\mu} \,,
\nonumber \\
\bar{\zeta}_{4}&=&-i\tilde{\sigma} \,,
\quad  \bar{\zeta}_{4,\theta}=-\tilde{\chi} \,,
\nonumber \\
\bar{\eta}&=&\varphi \,,
\quad \quad\:\,  \bar{\eta}_{,\theta}=-i\chi \,,
\label{32}
\end{eqnarray}
%
at $\lambda=\theta=0$, 
Eqs. (\ref{26}) and (\ref{27}) agree with Eqs. (\ref{1}) and (\ref{6}),
respectively.
Hence, the ordinary gauge theory of Abelian rank-2
antisymmetric tensor field is correctly reproduced
up to the gauge-fixing terms from the superspace formulation
characterized by the action $\bar{S}$.
In addition, the source term $\bar{S}_{\Sigma}$ and
the generating functional $W$ have the same forms
as those in the ordinary gauge theory
of Abelian rank-2 antisymmetric tensor field.
The generating functional $\bar{W}$ has a neat form, 
while $W$ is directly related to the ordinary gauge theory,
although $W$ is still a function of $\lambda$ and $\theta$.
These functional are related to each other by Eqs. (\ref{30}) and
(\ref{31}).

We now mention a difference between earlier superspace formulations
\cite{supe} and our superspace formulation:
in the earlier formulations, all the components of $\bar{B}_{ij}$
except $\bar{B}_{\mu\nu}$ are identified with the ghost fields
$\rho_{\mu}$, $\tilde{\rho}_{\mu}$, $\sigma$, $\varphi$
and $\tilde{\sigma}$.
On the other hand, in our formulation, only two components
$\bar{B}_{\mu5}$ and $\bar{B}_{55}$ are identified with
the ghost fields $\rho_{\mu}$ and $\sigma$,
while the other components except $\bar{B}_{\mu\nu}$
are treated as auxiliary fields.
Instead, $\bar{\zeta}_{\mu}$, $\bar{\zeta}_{4}$ and $\bar{\eta}$
are identified with ghost fields
$\tilde{\rho}_{\mu}$, $\tilde{\sigma}$ and $\varphi$, respectively.
As will be seen in the next section, the treatment of
$\bar{B}_{ij}$ in our formulation makes possible
to determine BRS transformation rules without assuming
the so-called horizontality condition.

\section{BRS transformation and six-dimensional gauge
transformation}
We shall see in this section that some of the BRS transformation rules
can be realized as the Euler-Lagrange equations and that a six-dimensional
gauge transformation is related to the BRS transformation.
Taking into account Eq. (\ref{32}), we define the
BRS transformation rules of the superfields so that
they can reduce to Eq. (\ref{2}):
\begin{eqnarray}
\mbox{\boldmath{$\delta$}}
\bar{B}_{\mu\nu} &=&
-i\partial_{\mu}\bar{B}_{\nu5}+i\partial_{\nu}\bar{B}_{\mu5}
\nonumber \\
\mbox{\boldmath{$\delta$}}
\bar{B}_{\mu5} &=& {i\over2}\partial_{\mu}\bar{B}_{55} \,,
\quad
\mbox{\boldmath{$\delta$}}\bar{B}_{55}=0 \,,
\;\;\;
\label{33}
\end{eqnarray}
%
and
\begin{eqnarray}
\mbox{\boldmath{$\delta$}}
\bar{\zeta}_{\mu}&=&i\bar{\zeta}_{\mu,\theta} \,,
\quad \quad \;\,
\mbox{\boldmath{$\delta$}}
\bar{\zeta}_{\mu,\theta}=0 \,,
\nonumber \\
\mbox{\boldmath{$\delta$}}
\bar{\zeta}_{4}&=&i\bar{\zeta}_{4,\theta} \,,
\quad \quad \;\,\,
\mbox{\boldmath{$\delta$}}
\bar{\zeta}_{4,\theta}=0 \,,
\nonumber \\
\mbox{\boldmath{$\delta$}}
\bar{\eta}&=&i\bar{\eta}_{,\theta} \,,
\quad \qquad \:
\mbox{\boldmath{$\delta$}}
\bar{\eta}_{,\theta}=0 \,.
\label{34}
\end{eqnarray}
%
The transformation rules (\ref{34}) indicate that the BRS
transformation $\mbox{\boldmath{$\delta$}}$ may be
represented as the derivative with respect to $\theta$.
If the equations
$\partial_{\mu}\bar{B}_{\nu5}
-\partial_{\nu}\bar{B}_{\mu5}=
-\bar{B}_{\mu\nu,\theta\,}$,
$\partial_{\mu}\bar{B}_{55}=2\bar{B}_{\mu5,\theta\,}$,
and $\bar{B}_{55,\theta}=0$ are satisfied, the BRS
transformation defined by Eqs. (\ref{33}) and (\ref{34})
is represented as
\begin{eqnarray}
\mbox{\boldmath{$\delta$}}=
i\frac{\partial}{\partial\theta} \,.
\label{35}
\end{eqnarray}
%
Remarkably, these desirable equations are derived from the action (\ref{18})
as the Euler-Lagrange equations for
$\bar{B}_{\mu\nu,\lambda}$, $\bar{B}_{\mu 4,\lambda}$ and
$\bar{B}_{44,\lambda\,}$:
\begin{eqnarray}
\frac{\partial\bar{S}}{\partial\bar{B}
\raisebox{0.95ex}{\mbox{\scriptsize{$\mu\nu$}}}
\raisebox{-0.36ex}{\mbox{\scriptsize{$,\!\lambda$}}}}
&=&-{1\over2}(\bar{B}_{\mu\nu,\theta}
+\partial_{\mu}\bar{B}_{\nu 5}-\partial_{\nu}\bar{B}_{\mu 5})=0 \,,
\label{36}
\\
\frac{\partial\bar{S}}{\partial\bar{B}
\raisebox{0.95ex}{\mbox{\scriptsize{$\mu$}}}
\raisebox{-0.36ex}{\mbox{\scriptsize{$4,\!\lambda$}}}}
&=&-2\bar{B}_{\mu 5,\theta}+\partial_{\mu}\bar{B}_{55}=0 \,,
\label{37}
\\
\frac{\partial\bar{S}}{\partial\bar{B}_{44,\lambda}}
&=&{3\over2}\bar{B}_{55,\theta}=0 \,.
\label{38}
\end{eqnarray}
%
Note here that the superfields are functions of $x^{\mu}$,
$\lambda$ and $\theta$, while $\bar{S}$ is a function of
$\lambda$ and $\theta$.
It should be emphasized that with Eq. (\ref{35}),
the BRS transformation rules (\ref{33}) can be obtained as the
Euler-Lagrange equations. This situation is quite different from
that in the earlier superspace formulations, in which the BRS
transformation rules are determined by putting in the horizontality
condition by hand.

We now extend the BRS transformation to the superfields
$\bar{B}_{\mu4}$, $\bar{B}_{44}$ and $\bar{B}_{45}$,
utilizing Eq. (\ref{35}) and the Euler-Lagrange equations
\begin{eqnarray}
\frac{\partial\bar{S}}{\partial\bar{B}
\raisebox{0.95ex}{\mbox{\scriptsize{$\mu$}}}
\raisebox{-0.36ex}{\mbox{\scriptsize{$5,\!\lambda$}}}}
&=&\bar{B}_{\mu 4,\theta}-\partial_{\mu}\bar{B}_{45}
+\bar{B}_{\mu 5,\lambda}=0 \,,
\label{39}
\\
\frac{\partial\bar{S}}{\partial\bar{B}_{55,\lambda}}
&=&{1\over2}\bar{B}_{44,\theta}+\bar{B}_{45,\lambda}=0 \,,
\label{40}
\\
\frac{\partial\bar{S}}{\partial\bar{B}_{45,\lambda}}
&=&-2\bar{B}_{45,\theta}-\bar{B}_{55,\lambda}=0 \,.
\label{41}
\end{eqnarray}
%
Since $\bar{S}_{\rm SG}$ in Eq. (\ref{18}) does not contain the
superfields $\bar{B}_{ij,\lambda}$, Eqs (\ref{36})-(\ref{41}) can
collectively be written as
\begin{eqnarray}
\frac{\partial\bar{S}_{0}}{\partial\bar{B}_{ij,\lambda}}=0 \,.
\label{42}
\end{eqnarray}
%
We can understand Eq. (\ref{42}) as the BRS transformation rule of
$\bar{B}_{ij}$.

Now, choosing the gauge parameter $\bar{\Lambda}_{i}$ in
Eq. (\ref{16}) to be a particular form
\begin{eqnarray}
\bar{\Lambda}_{i}(\bar{x})=\bar{B}_{i5}(\bar{x})\Lambda \,,
\label{43}
\end{eqnarray}
%
we define the six-dimensional gauge transformation
\begin{eqnarray}
\hat{\delta}\bar{B}_{ij}=
(\partial_{i}\bar{B}_{j5}
-(-1)^{|i||j|}\partial_{j}\bar{B}_{i5})\Lambda \,,
\label{44}
\end{eqnarray}
%
where $\Lambda$ is an anticommuting infinitesimal constant.
Using Eqs. (\ref{36})-(\ref{41}), we readily show that
\begin{eqnarray}
\hat{\delta}\bar{B}_{ij}
=\Lambda\bar{B}_{ij,\theta}
=\Lambda{{\partial\bar{B}_{ij}}\over{\partial\theta}} \,.
\label{45}
\end{eqnarray}
%
Differentiations of Eq. (\ref{45}) with respect to $\lambda$ and
$\theta$ lead to
\begin{eqnarray}
\hat{\delta}\bar{B}_{ij,\lambda}&=&\Lambda
{{\partial\bar{B}_{ij,\lambda}}\over{\partial\theta}} \,,
\\
\label{46}
\hat{\delta}\bar{B}_{ij,\theta}&=&\Lambda
{{\partial\bar{B}_{ij,\theta}}\over{\partial\theta}} \,. \;
\label{47}
\end{eqnarray}
%
As seen from Eqs. (\ref{35}) and (\ref{45})-(\ref{47}), 
the six-dimensional gauge transformation
$\hat{\delta}$ for
$\bar{B}_{ij}$, $\bar{B}_{ij,\lambda}$ and
$\bar{B}_{ij,\theta}$ is nothing other than 
the BRS transformation for them 
with the parameter $-i\Lambda$. 
Note that Eq.(\ref{47}) vanishes
because of $\partial^{2}/\partial\theta^{2}=0$ or
the nilpotency property of the BRS transformation.

\section{WT identity}
In this section we derive the WT identity for the gauge theory of
Abelian rank-2 antisymmetric tensor field by making use of the superspace
formulation discussed above. Since $\bar{S}_{0}$ is a functional
of the superfields $\bar{B}_{ij}$, $\bar{B}_{ij,\lambda}$ and
$\bar{B}_{ij,\theta}$, the use of Eqs.(\ref{45})-(\ref{47}) gives
\begin{eqnarray}
\hat{\delta} \bar{S}_{0}
=\Lambda
{{\partial\bar{S}_{0}}\over{\partial\theta}} \,.
\label{48}
\end{eqnarray}
%
Then, noting the invariance of $\bar{S}_{0}$ under the
gauge transformation (\ref{44}), we have
\begin{eqnarray}
{{\partial\bar{S}_{0}}\over{\partial\theta}}=0 \,,
\label{49}
\end{eqnarray}
%
which shows the BRS invariance of $\bar{S}_{0}$.
Differentiating Eq. (\ref{22}) with respect to $\theta$, we obtain
\begin{eqnarray}
{{\partial \bar{W}}\over{\partial\theta}}
&=& \int \{ {\cal D}\bar{B}_{ij} \}
\{ {\cal D}\bar{\zeta}_{i} \}
\{ {\cal D}\bar{\eta} \}
{{\partial}\over{\partial\theta}}
\exp(i\bar{S}) 
\nonumber \\
&=& \int \{ {\cal D}\bar{B}_{ij} \}
\{ {\cal D}\bar{\zeta}_{i} \}
\{ {\cal D}\bar{\eta} \}
i{{\partial\bar{S}_{0}}\over{\partial\theta}}
\exp(i\bar{S}) \,,
\label{50}
\end{eqnarray}
%
where the formula (\ref{14}) has been applied. 
From Eq. (\ref{49}), it follows that
\begin{eqnarray}
{{\partial \bar{W}}\over{\partial\theta}}=0 \,.
\label{51}
\end{eqnarray}
Instead of using the six-dimensional gauge invariance of $\bar{S}_{0}$,
we can show Eq. (\ref{51}) by directly calculating 
the last term of Eq. (\ref{50}).
In this alternative method,
field theoretical analogies of the formula $(x-y)\delta(x-y)=0$
are repeatedly used.

Differentiation of Eq. (\ref{31}) with respect to $\theta$ is
simply written as
\begin{equation}
{{\partial W}\over{\partial\theta}}
= \int {\cal D}\bar{K}^{4i}
{\cal D}\bar{K}\raisebox{0.97ex}{\mbox{\scriptsize{$4i$}}}
\raisebox{-0.36ex}{\mbox{\scriptsize{$,\!\theta$}}}
{\cal D}\bar{t}_{4} {\cal D}\bar{t}_{4,\theta}
{{\partial\bar{W}}\over{\partial\theta}}
\label{52}
\end{equation}
%
by taking into account
$\partial\bar{K}\raisebox{0.97ex}{\mbox{\scriptsize{$4i$}}}
\raisebox{-0.36ex}{\mbox{\scriptsize{$,\!\theta$}}}
/\partial\theta
=\partial\bar{t}_{4,\theta}/\partial\theta$=0,
and
$\int d\bar{K}\raisebox{0.97ex}{\mbox{\scriptsize{$4i$}}}
\raisebox{-0.36ex}{\mbox{\scriptsize{$,\!\theta$}}}
d\bar{K}\raisebox{0.97ex}{\mbox{\scriptsize{$4i$}}}
\raisebox{-0.36ex}{\mbox{\scriptsize{$,\!\theta$}}}
f_{1}(\bar{K}\raisebox{0.97ex}{\mbox{\scriptsize{$4i$}}}
\raisebox{-0.36ex}{\mbox{\scriptsize{$,\!\theta$}}})
=
\int d\bar{t}_{4,\theta}d\bar{t}_{4,\theta}
f_{2}(\bar{t}_{4,\theta})=0$
satisfied for arbitrary functions $f_{1}$ and $f_{2}$.
As a result, Eq. (\ref{51}) gives
\begin{eqnarray}
{{\partial W}\over{\partial\theta}}=0 \,.
\label{53}
\end{eqnarray}
%
On the other hand, calculating ${\partial W}/\partial\theta$ from
the expression (\ref{29}), we obtain
\begin{eqnarray}
{{\partial W}\over{\partial\theta}}
&=&
\int d^{4}x \left[\,
{1\over2}\bar{K}\raisebox{0.96ex}{\mbox{\scriptsize{$\mu\nu$}}}
\raisebox{-0.36ex}{\mbox{\scriptsize{$,\!\theta$}}}
{{\partial W}\over{\partial\bar{K}^{\mu\nu}}}
+ \bar{K}\raisebox{0.96ex}{\mbox{\scriptsize{$\mu5$}}}
\raisebox{-0.36ex}{\mbox{\scriptsize{$,\!\theta$}}}
{{\partial W}\over{\partial\bar{K}^{\mu5}}} \right.
\nonumber \\
& &\; \left.
+ \bar{t}\raisebox{0.95ex}{\mbox{\scriptsize{$\mu$}}}
\raisebox{-0.36ex}{\mbox{\scriptsize{$,\!\theta$}}}
{{\partial W}\over{\partial\bar{t}^{\mu}}}
+ \bar{t}_{5,\theta}
{{\partial W}\over{\partial\bar{t}_{5}}}
+ \bar{u}_{,\theta}
{{\partial W}\over{\partial\bar{u}}} \,\right]
\nonumber \\
&=&
-i\int d^{4}x \int{\cal D}\bar{\cal M} \left[\,
\bar{K}\raisebox{0.96ex}{\mbox{\scriptsize{$\mu\nu$}}}
\raisebox{-0.36ex}{\mbox{\scriptsize{$,\!\theta$}}}
(\partial_{\mu}\bar{B}_{\nu5}-\partial_{\nu}\bar{B}_{\mu5})
\right.
\nonumber \\
& &\;
\left.
+ \bar{K}\raisebox{0.96ex}{\mbox{\scriptsize{$\mu5$}}}
\raisebox{-0.36ex}{\mbox{\scriptsize{$,\!\theta$}}}
\partial_{\mu}\bar{B}_{55}
- \bar{t}\raisebox{0.95ex}{\mbox{\scriptsize{$\mu$}}}
\raisebox{-0.36ex}{\mbox{\scriptsize{$,\!\theta$}}}
\bar{\zeta}_{\mu,\theta}
+ \bar{t}_{5,\theta} \bar{\zeta}_{4,\theta}
+ \bar{u}_{,\theta} \bar{\eta}_{,\theta} \,\right]
\nonumber \\
& &\;\times
\exp(i\bar{S}') \;.
\label{54}
\end{eqnarray}
%
We now introduce sources $J^{\mu\nu}$, $J^{\mu}$, $j^{\mu}$,
$j$ and $\tilde{j}$ on $M^{4}$
that are defined, at $\lambda=\theta=0$, by
$J^{\mu\nu}=
-i\bar{K}\raisebox{0.96ex}{\mbox{\scriptsize{$\mu\nu$}}}
\raisebox{-0.36ex}{\mbox{\scriptsize{$,\!\theta$}}}$,
$J^{\mu}=
2i\bar{K}\raisebox{0.96ex}{\mbox{\scriptsize{$\mu5$}}}
\raisebox{-0.36ex}{\mbox{\scriptsize{$,\!\theta$}}}$,
$j^{\mu}=
\bar{t}\raisebox{0.95ex}{\mbox{\scriptsize{$\mu$}}}
\raisebox{-0.36ex}{\mbox{\scriptsize{$,\!\theta$}}}$,
$j=\bar{t}_{5,\theta}$ and
$\tilde{j}=i\bar{u}_{,\theta}$.
With these sources and Eq. (\ref{32}), it follows from
Eqs. (\ref{53}) and (\ref{54}) that
\begin{eqnarray}
& &\int d^{4}x \int{\cal D}{\cal M}
\left[ J^{\mu\nu}
(\partial_{\mu}\rho_{\nu}-\partial_{\nu}\rho_{\mu})
\right.
\nonumber \\
& & \;\left. +J^{\mu}\partial_{\mu}\sigma
+j^{\mu}\beta_{\mu}
+j\tilde{\chi}
+\tilde{j}\chi \right]=0 \,,
\label{55}
\end{eqnarray}
%
where
\begin{equation}
{\cal D}{\cal M}\equiv
{\cal D}B_{\mu\nu} {\cal D}\rho_{\mu} {\cal D}\sigma
{\cal D}\tilde{\rho}_{\mu} {\cal D}\tilde{\sigma}
{\cal D}\beta_{\mu} {\cal D}\tilde{\chi}
{\cal D}\varphi {\cal D}\chi \,.
\label{56}
\end{equation}
%
This is nothing other than the WT identity in the ordinary
gauge theory of Abelian rank-2 antisymmetric tensor field.
Hence, Eq. (\ref{53}) is understood as the WT identity.

Differentiating Eq. (\ref{30}) with respect to $\theta$, we have
\begin{equation}
{{\partial\bar{W}}\over{\partial\theta}}
= N\prod_{i,x} \delta(\bar{K}^{4i}(\bar{x}))
\delta(\bar{K}\raisebox{0.97ex}{\mbox{\scriptsize{$4i$}}}
\raisebox{-0.36ex}{\mbox{\scriptsize{$,\!\theta$}}}(\bar{x}))
\prod_{x} \delta(\bar{t}_{4}(\bar{x}))
\delta(\bar{t}_{4,\theta}(\bar{x}))
{{\partial W}\over{\partial\theta}} \,,
\label{57}
\end{equation}
%
where
$\partial \{ \delta(\bar{K}^{4i})
\delta(\bar{K}\raisebox{0.97ex}{\mbox{\scriptsize{$4i$}}}
\raisebox{-0.36ex}{\mbox{\scriptsize{$,\!\theta$}}})\}/
\partial\theta
=
\delta'(\bar{K}^{4i})
\bar{K}\raisebox{0.97ex}{\mbox{\scriptsize{$4i$}}}
\raisebox{-0.36ex}{\mbox{\scriptsize{$,\!\theta$}}}
\delta(\bar{K}\raisebox{0.97ex}{\mbox{\scriptsize{$4i$}}}
\raisebox{-0.36ex}{\mbox{\scriptsize{$,\!\theta$}}})=0$
and
$ \partial\{ \delta(\bar{t}_{4})
\delta(\bar{t}_{4,\theta}) \}/
\partial\theta
=\delta'(\bar{t}_{4})
\bar{t}_{4,\theta}\delta(\bar{t}_{4,\theta})=0$ have been used. 
From Eqs. (\ref{52}) and (\ref{57}), we see that Eq. (\ref{51}) 
is equivalent to
Eq. (\ref{53}). Therefore Eq. (\ref{51}) is considered the WT identity
written in terms of the superspace formulation.

\section{summary and discussion}

In this paper, we have found that the superspace formulation 
proposed by Joglekar works well not only in the Yang-Mills 
theory but also in the gauge theory of Abelian rank-2 antisymmetric 
tensor field. As we have seen, all information on the quantization of 
an Abelian rank-2 antisymmetric tensor field is contained in the 
simple generating functional $\bar{W}$ in Eq. (\ref{22}), with which 
the WT identity is expressed in a compact and elegant form 
\begin{eqnarray}
{{\partial \bar{W}}\over{\partial\theta}}=0 \,.
\label{58}
\end{eqnarray}
%

In the superspace formulation of Yang-Mills theory, 
the WT identities are also cast in 
the same form as Eq. (\ref{58}) \cite{sdj2}. 
It was shown in Ref.\cite{j&m1} that this simple form can 
directly be derived from partial $\rm{OSp}(3,1|2)$ invariance 
of the generating functional $\bar{W}$ 
defined in the superspace formulation of Yang-Mills theory. 
Furthermore, several subjects concerning the renormalization 
of Yang-Mills theory \cite{j&m2} and generalizations of 
the BRS transformation \cite{j&m3} have been studied  
in the context of the superspace formulation of Yang-Mills theory. 
Based on the superspace formulation considered in the present paper, 
we will be able to extend the discussions in Ref.[15-17] to 
the gauge theory of Abelian rank-2 antisymmetric tensor field.

The superspace formulation can readily be applied to the 
gauge theories of Abelian antisymmetric tensor fields of higher rank. 
It is also possible to generalize the superspace formulation 
in the present paper to several of the gauge theories containing 
both the Yang-Mills and rank-2 antisymmetric tensor fields. 
One of such theories is the theory involving the Chapline-Manton 
coupling \cite{Chap} that is defined by the action 
\begin{equation}
S_{\rm CM}= \int d^{4}x \left[ 
-{1\over4} {F_{\mu\nu}}^{a} F^{\mu\nu a} 
+{1\over12} H_{\mu\nu\rho}H^{\mu\nu\rho} \right] \,, 
\label{59}
\end{equation}
%
with $H_{\mu\nu\rho}\equiv \partial_{\mu}B_{\nu\rho}+
\partial_{\nu}B_{\rho\mu}+\partial_{\rho}B_{\mu\nu}
+k\omega_{\mu\nu\rho}$. 
Here ${F_{\mu\nu}}^{a}$ is the field strength of 
Yang-Mills fields ${A_{\mu}}^{a}$, $k$ a constant with dimensions 
of length, and $\omega_{\mu\nu\rho}$  
the Chern-Simons three-form consisting of ${A_{\mu}}^{a}$. 
Another theory that one thinks is a massive gauge theory of 
non-Abelian rank-2 antisymmetric tensor field 
\cite{d&n} defined  by the action 
\begin{equation}
S_{\rm NA}= \int d^{4}x \left[ 
-{1\over4} {F_{\mu\nu}}^{a} F^{\mu\nu a} 
+{1\over12} 
\hat{H}_{\mu\nu\rho}\raisebox{0.98ex}{\mbox{\scriptsize{$a$}}}
\hat{H}^{\mu\nu\rho a} 
-{1\over4}m^{2}
\hat{B}_{\mu\nu}\raisebox{0.98ex}{\mbox{\scriptsize{$a$}}}
\hat{B}^{\mu\nu a} 
\right] \,, 
\label{60}
\end{equation}
%
with 
$\hat{B}_{\mu\nu}\raisebox{0.98ex}{\mbox{\scriptsize{$\!a$}}}
\equiv {B_{\mu\nu}}^{a}-m^{-1}
(D_{\mu}{\phi_{\nu}}^{a}-D_{\nu}{\phi_{\mu}}^{a})$ 
and 
$\hat{H}_{\mu\nu\rho}\raisebox{0.98ex}{\mbox{\scriptsize{$a$}}}
\equiv
D_{\mu}\hat{B}_{\nu\rho}\raisebox{0.98ex}{\mbox{\scriptsize{$a$}}}
+D_{\nu}\hat{B}_{\rho\mu}\raisebox{0.98ex}{\mbox{\scriptsize{$a$}}}
+D_{\rho}\hat{B}_{\mu\nu}\raisebox{0.98ex}{\mbox{\scriptsize{$a$}}}$. 
Here ${{B}_{\mu\nu}}^{a}$ is a non-Abelian antisymmetric 
tensor field, ${\phi_{\mu}}^{a}$ a non-Abelian vector 
field, $m$ a constant with dimensions of mass, and  
$D_{\mu}$ denotes the covariant derivative defined from 
${A_{\mu}}^{a}$. 
In addition to the theories defined by the Lagrangians 
(\ref{59}) and (\ref{60}), 
there are gauge theories 
with topological terms consisting of the Yang-Mills and 
antisymmetric tensor fields [20,21]. 
To apply the superspace formulation proposed by Joglekar  
to those theories, it is necessary to consider how topological terms 
are defined in the superspace.

\begin{acknowledgements}
We are grateful to Profs. S. Ishida, S. Naka and 
other members of the Theoretical 
Physics Group at Nihon University for their encouragements. 
One of us (B.P.M.) acknowledges the hospitality and financial support 
provided by Atomic Energy Research Institute, Nihon University. 
B.P.M. is also grateful
to Prof. Satish D. Joglekar of IIT-Kanpur, India for encouragements and
many fruitful discussions.
This work was supported in part by Nihon University Research Grant. 
\end{acknowledgements}


\begin{references}
\bibitem{kalb} M. Kalb and P. Ramond, Phys. Rev. D {\bf 9}, 2273 (1974).
\bibitem{lund} F. Lund and T. Regge, Phys. Rev. D {\bf 14}, 1524 (1976);
M. Sato and S. Yahikozawa, Nucl. Phys. {\bf B436}, 100 (1995).
\bibitem{suga} A. Sugamoto, Phys. Rev. D {\bf 19}, 1820 (1979);
R. L. Davis and E. P. S. Shellard, Phys. Lett. B {\bf 214}, 219 (1988).
\bibitem{sgr} A. Salam and E. Sezgin,
{\it Supergravities in Diverse Dimensions}
(North-Holland, Amsterdam and World Scientific, Singapore, 1989).
\bibitem{str} M. B. Green, J. H. Schwarz, and E. Witten,
{\it Superstring Theory}
(Cambridge University Press, New York, 1987);
J. Polchinski, {\it String Theory}
(Cambridge University Press, New York, 1998);
\bibitem{top} E. Cremmer and J. Scherk, Nucl. Phys. {\bf B72}, 117 (1974);
A. Aurilia and Y. Takahashi, Prog. Theor. Phys. {\bf 66}, 693 (1981);
I. Oda and S. Yahikozawa, {\it ibid.} {\bf 83}, 991 (1990);
T. J. Allen, M. J. Bowick and A. Lahiri, Mod. Phys. Lett. A {\bf 6},
559 (1991);
S. Deguchi, T. Mukai and T. Nakajima, Phys. Rev. D59,
65003 {\bf 59} (1999).
\bibitem{loop} P. G. O. Freund and R. I. Nepomechie,
Nucl. Phys. {\bf B199}, 482 (1982);
J. A. de Azc\'{a}rraga, J. M. Izquierdo, and P. K. Townsend,
Phys. Rev. D {\bf 45}, R3321 (1992);
S. Deguchi and T. Nakajima, Int. J. Mod. Phys. A {\bf 9},
1889 (1994).
\bibitem{town} P. K. Townsend, Phys. Lett. {\bf 88B}, 97 (1979).
\bibitem{kimu} T. Kimura, Prog. Theor. Phys. {\bf 64}, 357 (1980);
H. Hata, T. Kugo, and N. Ohta,
Nucl. Phys. {\bf B178}, 527 (1981);
M. Henneaux and C. Teitelboim,
{\it Quantization of Gauge Systems}
(Princeton University Press, Princeton, 1992);
J. Gomis, J. Paris, and S. Samuel, Phys. Rep. {\bf 259},
1 (1995).
\bibitem{supe} J. Thierry-Mieg and L. Baulieu, Nucl. Phys.
{\bf B228}, 259 (1983);
J. Barcelos-Neto and R. Thibes, J. of Math. Phys {\bf 39},
5669 (1998).
%
\bibitem{sdj} S. D. Joglekar (unpublished);
Phys. Rev. D {\bf 43}, 1307 (1991); {\bf 48}, 1878(E) (1993). 
\bibitem{ssp}
L. Bonora and M. Tonin, Phys. Lett {\bf 98 B}, 48 (1981);
A. C. Hirshfeld and H. Leschke, Phys. Lett. {\bf 101 B}, 48, (1981);
R. Delbourgo and P. D. Jarvis, J. Phys. A {\bf 15}, 611 (1982);
L. Baulieu and J. Thierry-Mieg, Nucl. Phys. {\bf B197}, 477 (1982);
L. Baulieu, Phys. Rep. {\bf 129}, 1 (1985).
\bibitem{bpm} B. P. Mandal,
^^ ^^ {\it BRS symmetry and superspace formulation of
gauge theories}", PhD thesis, IIT-Kanpur (1996).
\bibitem{sdj2} S. D. Joglekar; 
Phys. Rev. D {\bf 44}, 3879 (1991); {\bf 48}, 1879(E) (1993). 
\bibitem{j&m1} S. D. Joglekar and B. P. Mandal; 
Phys. Rev. D {\bf 49}, 5382 (1994). 
\bibitem{j&m2} S. D. Joglekar and B. P. Mandal; 
Phys. Rev. D {\bf 49}, 5617 (1994); 
{\bf 55}, 5038 (1997). 
\bibitem{j&m3} S. D. Joglekar and B. P. Mandal; 
Z. Phys. C {\bf 70}, 673 (1996); {\bf 74}, 179 (1997).
\bibitem{Chap} G. F. Chapline and N. S. Manton, 
Phys. Lett. {\bf 120B}, 105 (1983); 
S. Deguchi and T. Nakajima, Mod. Phys. Lett. A 
{\bf 12}, 111 (1997).  
\bibitem{d&n} S. Deguchi and T. Nakajima, 
Int. J. Mod. Phys. A {\bf 10}, 1019 (1995). 
\bibitem{top2}
D. Birmingham, M. Blau, M. Rakowski and G. Thompson, 
Phys. Rep. {\bf 209}, 129 (1991); 
O. Piguet and S. P. Sorella, 
{\it Algebraic Renormalization} (Springer, Heidelberg, 1995). 
\bibitem{top3} 
D. Z. Freedman and P. K. Townsend, Nucl. Phys. {\bf B177}, 282 (1981); 
A. Lahiri, ^^ ^^ Generating vector boson masses", hep-th/9301060; 
Phys. Rev. {\bf D55}, 5045 (1997); 
J. Barcelos-Neto, A. Cabo and M. B. D. Silva, Z. Phys. 
{\bf C72}, 345 (1996); 
S. Deguchi, ^^ ^^ A universal Lagrangian for massive 
Yang-Mills theories without Higgs bosons", hep-th/9903135, 
NUP-A-98-8.  
 



\end{references}
\end{document}